\documentclass[aps,pra,showpacs]{revtex4}
\usepackage{amsmath}
\usepackage{graphicx}
\usepackage{dcolumn}
\usepackage{longtable}

\newcommand{\be}{\begin{equation}}
\newcommand{\ee}{\end{equation}}
\newcommand{\bearray}{\begin{eqnarray}}
\newcommand{\eearray}{\end{eqnarray}}
\newcommand{\bse}{\begin{subequations}}
\newcommand{\ese}{\end{subequations}}
 
\newcommand{\half}{\frac{1}{2}}

\newcommand{\fourth}{\frac{1}{4}}

\begin{document}

\title{Positronium energy levels at order $m \alpha^7$: Product contributions in the two-photon-annihilation channel.}

\author{Gregory S. Adkins}
\email[]{gadkins@fandm.edu}
\author{Lam M. Tran}
\author{Ruihan Wang}
\affiliation{Franklin \& Marshall College, Lancaster, Pennsylvania 17604}

\date{\today}

\begin{abstract}
Ongoing improvements in the measurement of positronium transition intervals motivate the calculation of the $O(m \alpha^7)$ corrections to these intervals.  In this work we focus on corrections to the spin-singlet parapositronium energies involving virtual annihilation to two photons in an intermediate state.  We have evaluated all contributions to the positronium S-state energy levels that can be written as the product of a one-loop correction on one side of the annihilation event and another one-loop correction on the other side.  These effects contribute $\Delta E = -0.561971(25) m \alpha^7/\pi^3$ to the parapositronium ground state energy.
\end{abstract}



\maketitle


\section{Introduction}
\label{introduction}

Positronium, the exotic atom composed of an electron and its antimatter partner the positron, plays a crucial part in our understanding of binding in quantum field theory.  Positronium has several interesting properties that, taken together, set it apart from other Coulombic two-body bound states.  The constituents of positronium are pointlike particles with no complicating internal structure.  The dynamics of positronium is governed almost completely by QED--strong and electromagnetic effects are negligible at current levels of precision.  Recoil effects in positronium are maximal--the mass ratio of the constituents is one.  Finally, positronium is subject to real and virtual annihilation into photons.  Positronium is accessible to high precision measurements of energy levels, decay rates, branching ratios, etc., and so forms an ideal system for testing relevant theories, calculational methods, and experimental techniques.

Positronium was first produced in 1951 by Deutsch [1], who with Dulit made the first measurement of a transition energy--the ground state hyperfine splitting (hfs) [2].  (Details about the discovery of positronium along with reflections and background material are given in \cite{Maglic75}.)  Numerous measurements of the ground state hfs, the $n=2$ fine structure intervals, and the 2S-1S interval were made over the years as summarized in a number of reviews \cite{Debenedetti54,Hughes73,Berko80,Rich81,Mills90,Karshenboim04,Karshenboim05,Namba12}.  The highest precision results for the $n=1$ hfs \cite{Mills75,Mills83,Ritter84,Ishida14}, $n=2$ fine structure \cite{Mills75b,Hatamian87,Hagena93,Ley94}, and 2S-1S interval \cite{Fee93} all have uncertainties near the $1 MHz$ level.  Recently there have been vigorous efforts to develop new approaches to the measurement of the various energy intervals and to improve their precision \cite{Fan96,Ley02,Cassidy08,Sasaki11,Crivelli11,Ishida12,Yamazaki12,Namba12,Cassidy12,Mills14,Cooke15,Miyazaki15}, raising the promise of significant progress in the not-too-distant future.

The corrections to the Bohr energy levels of positronium can be expressed as a double series in $\alpha$ and $L \equiv \ln \left ( 1/\alpha \right )$:
\be \label{ps_th}
\Delta E = m \alpha^4 \Bigl \{ C_0 + +C_{11} \alpha L + C_{10} \alpha + C_{21} \alpha^2 L + C_{20} \alpha^2 + C_{32} \alpha^3 L^2 + C_{31} \alpha^3 L + C_{30} \alpha^3 + \cdots \Big \} .
\ee
All positronium energies through terms of order $m \alpha^6=18.7 MHz$ are known \cite{Elkhovsky94,Pachucki98,Czarnecki99a,Zatorski08} along with the leading
log correction of order $m \alpha^7 L^2 = 3.30 MHz$ \cite{Karshenboim93,Melnikov99,Pachucki99}.  The subleading log correction of order $m \alpha^7 L = 0.67 MHz$ is known for the hfs \cite{Kniehl00,Melnikov01,Hill01}.  Several of the pure order $m \alpha^7 = 0.14 MHz$ corrections are known as well.  Some, those involving ``ultrasoft'' photons (those having energy and momentum of order $m \alpha^2$) give contributions as large as several tenths of a $MHz$ \cite{Marcu11,Baker14}.  Corrections at $O(m \alpha^7)$ can be classified as either annihilation (involving intermediate states consisting only of photons) or non-annihilation, and the annihilation contributions can be organized by the number of intermediate-state photons.  Some non-annihilation corrections have been calculated \cite{Marcu11,Adkins14a,Eides1415}.  All terms having a one-photon intermediate state are known \cite{Baker14}, as are all terms having a three-photon intermediate state \cite{Adkins15c}.  Terms involving a two-photon intermediate state are the subject of this paper.

Contributions to the parapositronium energies at $O(m \alpha^7)$ that involve virtual annihilation to two photons can be organized into four groups.  Contributions that contain a light-by-light scattering subgraph as in Fig.~1(a) have been evaluated \cite{Adkins14b} as have those involving vacuum polarization corrections to the annihilation photons as in Fig.~1(b) \cite{Adkins15a}.   In the present work we give the result for all contributions containing one-loop corrections on either side of the virtual annihilation process as in Fig.~1(c).  We call these terms the ``product'' contribution as they involve the product of two one-loop self energy, vertex, or ladder corrections.  Terms that involve two-loop corrections either before or after the virtual annihilation as in Fig~1(d) remain to be done.

This work is organized as follows.  In Sec.~\ref{lower} we give some details of the calculation of the $O(m \alpha^5)$ and $O(m \alpha^6)$ two-photon-annihilation contributions.  Important aspects of the calculational procedure and notation are introduced in this section.  In Sec.~\ref{product} we describe our calculation of the $O(m \alpha^7)$ product terms.  Sec.~\ref{results} contains our results.

\begin{figure}[t]
\includegraphics[width=5.2in]{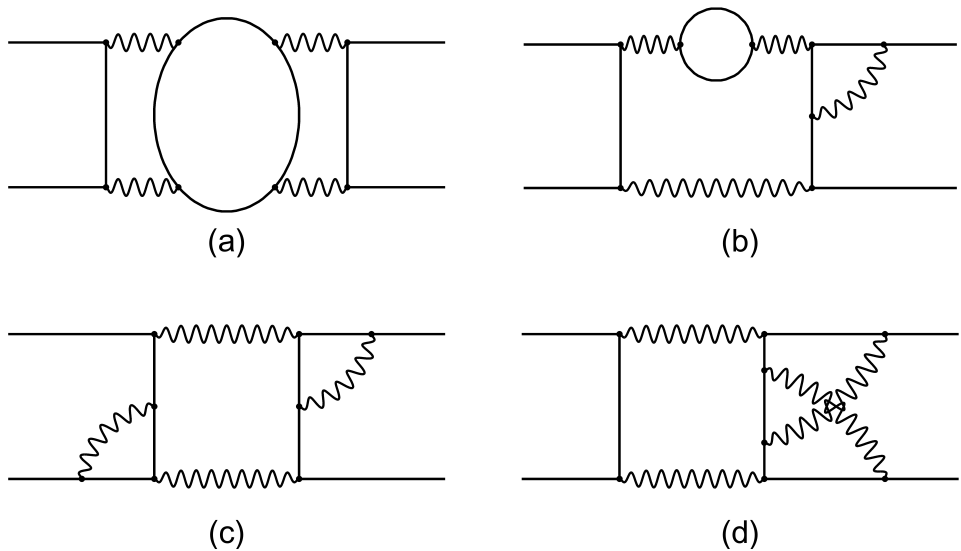}
\caption{\label{fig1} Representatives of the four classes of two-photon-annihilation corrections that contribute to the parapositronium energy levels at $O(m \alpha^7)$.  They are (a) light-by-light corrections, (b) terms involving vacuum polarization corrections to the annihilation photons, (c) terms containing the product of two one-loop corrections, one on each side of the virtual annihilation, and (d) terms with a two-loop correction on one side of the virtual annihilation. }
\end{figure}


\section{Lower Order Contributions}
\label{lower}

In this section we evaluate the two-photon-annihilation ($2 \gamma \mathrm{A}$) contributions at orders $m \alpha^5$ (done originally by Karplus and Klein \cite{Karplus52}) and $m \alpha^6$ (see \cite{Cung78,Adkins93}) in order to illustrate our calculational method and set notation.  We are free to use any convenient bound state formalism because, for the calculation at hand, the energy levels depend only on the spin structure of the wave function and its value at spatial contact but not on more detailed aspects of the formalism.  We use the formalism of \cite{Adkins99} in which the energy shift is the expectation value
\be 
\Delta E = i \bar{\Psi} \delta K \Psi .
\ee
of a two-particle-irreducible interaction kernel taken between appropriate wave functions.  The positronium states can be approximated as
\be
\Psi \rightarrow \phi_0 \begin{pmatrix} 0 & \chi \\ 0 & 0 \end{pmatrix}
\quad \bar \Psi^T \rightarrow \phi_0 \begin{pmatrix} 0 & 0 \\ \chi^\dagger & 0 \end{pmatrix} ,
\ee
where $\chi=1/\sqrt 2$ is the spin-singlet $2 \times 2$ matrix and $\phi_0=\sqrt{m^3 \alpha^3/(8 \pi n^3)}$ is the wave function at spatial contact for a state of principal quantum number $n$ and orbital angular momentum $\ell=0$.

\begin{figure}[t]
\includegraphics[width=4.5in]{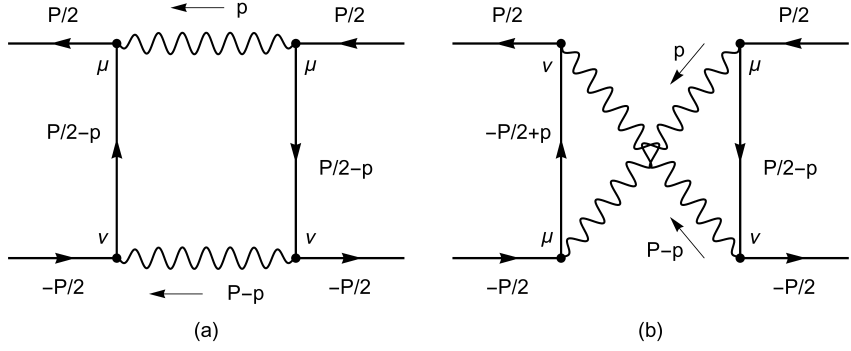}
\caption{\label{fig2} The LO $2 \gamma \mathrm{A}$ contributions.  Both (a) uncrossed and (b) crossed photons are included to account for Bose symmetry.  The wave functions bracketing these kernels on the left and right are implicit.}
\end{figure}

The graphs making leading-order (LO) $2 \gamma \mathrm{A}$ energy contributions of $O(m \alpha^5)$ are pictured in Fig.~\ref{fig2}.  The energy shift due to the graph of Fig.~\ref{fig2}(a) is
\bearray \label{energy_1a_explicit}
\Delta E_{a} &=& (-1) i \phi_0^2 \int \frac{d^4 p}{(2 \pi)^4}  \frac{-i}{p^2} \frac{-i}{(P-p)^2} 
\; \mathrm{tr} \Bigl [ \begin{pmatrix} 0 & 0 \\ \chi^\dagger & 0 \end{pmatrix} (-i e \gamma^\mu) \frac{i}{\gamma (P/2-p)-m} (-i e \gamma^\nu)   \Bigr ] \cr
&\times& \mathrm{tr} \Bigl [ (-i e \gamma_\nu) \frac{i}{\gamma (P/2-p)-m} (-i e \gamma_\mu) \begin{pmatrix} 0 & \chi \\ 0 & 0 \end{pmatrix} \Bigr ]
\eearray
where $P = (2m,0)$ is the approximate positronium 4-momentum in the center-of-mass frame and the initial $(-1)$ is a fermionic sign factor required for annihilation graphs.  We can simplify this expression by using projection operators to write the spin matrices in terms of gamma matrices (in the Dirac representation) \cite{Itzykson80}:
\bse
\bearray
\begin{pmatrix} 0 & \chi \\ 0 & 0 \end{pmatrix} &=& \frac{1}{2 \sqrt{2}} \left ( \gamma n + 1 \right ) \gamma_5 , \\
\begin{pmatrix} 0 & 0 \\ \chi^\dagger & 0 \end{pmatrix} &=& \frac{1}{2 \sqrt{2}} \left ( - \gamma n + 1 \right ) \gamma_5 
\eearray
\ese
where $n=(1,\vec 0 \,)$ is the timelike unit vector.  We also scale a factor of the electron mass $m$ out of each momentum vector and give the dimensionless momentum $p/m$ the same name $p$ as before.  The LO contribution is
\be
\Delta E_{LO} = I_\mathrm{LO} \frac{m \alpha^5}{\pi}
\ee
where
\be \label{lowest_order_1}
I_\mathrm{LO} = 4 \int_0^\infty d \vert \vec{p} \, \vert \,  \int \frac{d p_0}{2 \pi i} \frac{\vec p \, ^2 T}{p^2 (p-2n)^2}
\ee
with
\be \label{lowest_order_traces}
T \equiv \fourth \mathrm{tr} \Bigl [\gamma_\nu \left [ \gamma(n-p)-1 \right ]^{-1} \gamma_\mu (\gamma n+1) \gamma_5 \Bigr ]
\fourth \mathrm{tr} \Bigl [\gamma^\mu \left [ \gamma(n-p)-1 \right ]^{-1} \gamma^\nu (-\gamma n+1) \gamma_5 \Bigr ].
\ee
It is easy to see that the contributions of both parts of Fig.~\ref{fig2} are equal, and the factor of 2 has been included in (\ref{lowest_order_1}).  The $\vert \vec p \, \vert$ integral is over the magnitude of the 3-vector $\vec p$ and runs from 0 to $\infty$.  The trace evaluates to $T=-2 \vec p\,^2/((n-p)^2-1)^2$, and so
\be
I_\mathrm{LO} = -8 \int_0^\infty dp \, p^4 \int \frac{d p_0}{2 \pi i} \left \{ (p_0^2-p^2+i \epsilon)[(p_0-2)^2-p^2+i \epsilon] [(p_0-1)^2-\omega_p^2+i \epsilon]^2 \right \}^{-1}
\ee
where now $p$ stands for $\vert \vec p \, \vert$, $\omega_p \equiv (p^2+1)^{1/2}$, and we have reinserted the $i \epsilon$ factors that were implicit up until now.  We use the residue theorem to evaluate the $p_0$ integral after closing the contour with an infinite semicircle in either the upper or lower half plane, yielding a result of the form $A(p)/(p-1-i \epsilon)+B(p)$.  The singularity at $p=1$ represents the possibility that the annihilation photons could be real \cite{Cutkosky60}, and leads to an imaginary part of $\Delta E$ that is connected to the parapositronium decay rate according to
\be \label{rate_imE}
\Gamma = -2 \mathrm{Im} \Delta E .
\ee
We extract the imaginary part using the identity
\bearray
\int_0^\infty dp \, \frac{A(p)}{p-1-i \epsilon} &=& \int_0^2 dp \left ( \frac{A(1)}{p-1-i \epsilon} + \frac{A(p)-A(1)}{p-1} \right ) + \int_2^\infty dp \frac{A(p)}{p-1} \cr
&=& i \pi A(1) + \int_0^\infty dp \frac{A(\tilde p)}{p-1} ,
\eearray
where
\be
A(\tilde p) = \begin{cases} A(p)-A(1) & \text{if} \; \; 0<p<2 \cr A(p) & \text{if} \; \; 2<p \end{cases}
\ee
as in earlier related calculations \cite{Adkins01a,Adkins14b}.  The LO energy shift is
\be
I_\mathrm{LO} = \half \ln 2 - \half - \frac{i \pi}{4}
\ee
in units of $m \alpha^5/\pi$.

\begin{figure}
\includegraphics[width=6.2in]{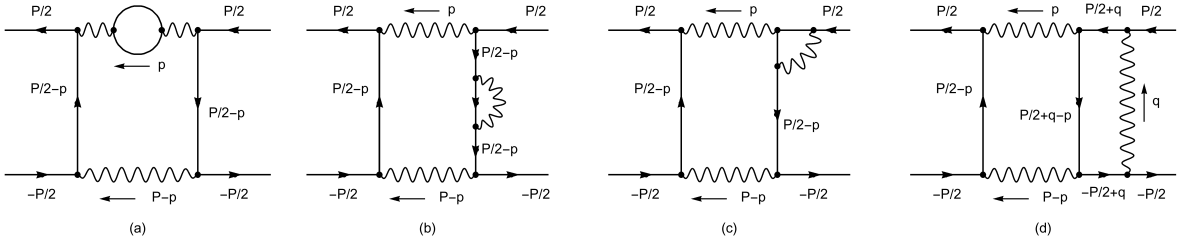}
\caption{\label{fig3} The four types of one-loop radiative corrections in the $2 \gamma \mathrm{A}$ channel that contribute at NLO:  vacuum polarization (a), self energy (b), vertex (c), and ladder (d).  Additional contributions having crossed annihilation photons and with the radiative corrections acting on different parts of the graphs give contributions equal to those shown and are included by means of appropriate symmetry factors.}
\end{figure}

The next-to-leading-order (NLO) graphs that contribute $O(\alpha)$ corrections to $\Delta E_{LO}$ are shown in Fig.~\ref{fig3}.  Each of these graphs comes with a symmetry factor giving the number of identical contributions represented by the graph.  For the vacuum polarization (VP) contribution of Fig.~\ref{fig3}(a), the symmetry factor is 4, coming from the two photons that the vacuum polarization correction could act on and the two configurations of annihilation photons: uncrossed and crossed.  The symmetry factors for the self energy (SE) of Fig.~\ref{fig3}(b), the vertex (V) of Fig.~\ref{fig3}(c), and the ladder (lad) of Fig.~\ref{fig3}(d) corrections are 4, 8, and 4, respectively.

Our approach is to start with the LO expression (\ref{lowest_order_1}) and insert a vacuum polarization, self energy, or vertex correction or a ladder photon.  We give our $O(\alpha)$ corrections in units of $m \alpha^6/\pi^2$.  The vacuum polarization correction has the value \cite{Adkins93,Adkins15a}
\be
I_{VP} = -\frac{1}{6} \zeta(2).
\ee
We used the Feynman gauge versions of the one-loop self energy and vertex parts from \cite{Adkins01b}.  The renormalized electron self energy function corrects the bare electron propagator of 4-momentum $p$ according to
\be
\frac{1}{\gamma p - 1} \rightarrow \frac{\alpha}{\pi} \Bigl [ S_1 + S_2(p) \Bigr ] \frac{1}{\gamma p-1}
\ee
where
\be \label{S1_def}
S_1 = \ln \lambda + \half
\ee
and
\be
S_2 = \int dx du f_{SE}(u) \frac{N(p)}{D(p)}
\ee
with $f_{SE}(u) = -1/(2 u)$, $N(p)=\{ 2-(1-x) \gamma p \} (\gamma p+1)$, and $D(p)=p^2-1-x/[(1-x)u]$.  A photon mass $\lambda$ was introduced in the course of mass-shell renormalization to regularize the infrared divergence, and all parametric integrals run from 0 to 1.  The renormalized vertex function corrects the bare vertex $\gamma^\mu$ according to the replacement
\be
\gamma^\mu \rightarrow \frac{\alpha}{\pi} \Bigl \{ V_1^\mu + V_2^\mu(p',p) + V_3^\mu(p',p) \Bigr \}
\ee
where
\bse
\bearray
V_1^\mu &=& \gamma^\mu \left ( -\ln \lambda - \frac{5}{4} \right ), \\
V_2^\mu(p',p) &=& \int dx du \frac{-N^\mu}{4 H}, \\
V_3^\mu(p',p) &=& \gamma^\mu \int dx du dz \frac{-x(H-x)}{2 \overline{H}},
\eearray
\ese
with
\bse
\bearray
N^\mu &=& \gamma^\lambda \left [ \gamma(p'+Q)+1 \right ] \gamma^\mu \left [ \gamma(p+Q) +1 \right ] \gamma_\lambda, \\
H &=& (1-x) \left [ u(1-p'^2)+(1-u) (1-p^2) \right ] - x u (1-u) (p'-p)^2 + x , \\
\overline{H} &=& x + z(H-x) ,
\eearray
\ese
where $p$ and $p'$ are the incoming and outgoing momenta (as in Fig.~1c of \cite{Adkins01b}) and $Q = -x [ u p' + (1-u) p ]$.  Finally, the ladder correction as applied to, say, the right hand side of (\ref{lowest_order_traces}) has the form:
\be \label{ladder_correction_1}
\gamma^\mu \frac{1}{\gamma (n-p)-1} \gamma^\nu \rightarrow \int \frac{d^4 q}{(2 \pi)^4} (-i e \gamma^\kappa) \frac{i}{\gamma(-n+q)-1} \gamma^\mu \frac{1}{\gamma (n+q-p)-1} \gamma^\nu \frac{i}{\gamma(n+q)-1} (-i e \gamma_\kappa) \frac{-i}{q^2-\lambda^2} .
\ee

Using the techniques described above, the one-loop self energy and vertex corrections to the parapositronium energies are straightforward to obtain.  The results are shown in Table~\ref{table1}.  Separate contributions for each part of the self energy and vertex functions are shown along with the totals.  The ladder contribution from (\ref{ladder_correction_1}) is a bit more involved, having the form
\be \label{ladder_correction_2}
F(p) \equiv -i (4 \pi \alpha)^2 \int \frac{d^4 q}{(2 \pi)^4} \frac{N(p,q)}{D(q) Z(p,q)}
\ee
where 
\bse
\bearray
D(q) &=& [(n-q)^2-1] [(n+q)^2-1] (q^2-\lambda^2) , \\
Z(p,q) &=& [(n+q-p)^2-1] , \\
N(p,q) &=& \gamma^\kappa [\gamma(-n+q)+1] \gamma^\mu [\gamma(n+q-p)+1] \gamma^\nu [\gamma(n+q)+1] \gamma_\kappa .
\eearray
\ese
The ladder correction has a binding singularity that is regulated by the photon mass.  This singularity is present because we are working just at the threshold for real production of the electron-positron pair and have set the relative momentum to zero and the positronium mass to $2m$.  The singularity can be isolated by writing
\be \label{ladder_splitup}
\frac{N(p,q)}{D(q) Z(p,q)} = \frac{N(p,0)}{D(q) Z(p,0)} + \frac{N(p,0)}{D(q)} \left ( \frac{1}{Z(p,q)}-\frac{1}{Z(p,0)} \right ) + \frac{N(p,q)-N(p,0)}{D(q) Z(p,q)} .
\ee
We label these terms 1, 2, and 3.  The first contains the singularity and its trace factor is proportional to the LO trace:
\be
\mathrm{tr} \Big [ N(p,0) (\gamma n+1) \gamma_5 \Big ] = -4 \mathrm{tr} \Big [\gamma^\mu [\gamma(n-p)+1] \gamma^\nu (\gamma n+1) \gamma_5 \Big ].
\ee
In this term the whole $q$ dependence is isolated in $D(q)$, and the integral of $1/D(q)$ is given by \cite{Adkins02}
\be
\int \frac{d^4 q}{i \pi^2} \frac{-1}{D(q)} = \frac{\pi}{\lambda} + \ln \lambda - 1 + O(\lambda) .
\ee
The linear binding singularity is the photon mass regularization scheme version of the ``Sommerfeld Factor'' that represents the effect of Coulombic binding and should be removed from the radiative correction \cite{Harris57,Alekseev59,Stroscio74,Caswell77}.  It can also be interpreted as an infrared effect that is cancelled in the course of the matching procedure in the context of an effective non-relativistic field theory approach to the problem \cite{Caswell86,Labelle94,Adkins02}.  The net result is that the ladder correction acts as the sum of three terms, $(\alpha/\pi) \left ( L_1+L_2+L_3 \right )$, where 
\be \label{L1_def}
L_1 = \ln \lambda -1
\ee
and multiplies the LO contribution, while $L_2$ and $L_3$ come from the final two terms of (\ref{ladder_splitup}).

The values for the NLO energy corrections are shown in Table~\ref{table1}.  We note that the total infrared divergence vanishes, as it must.  The real parts for SE2, V2, V3, lad2, and lad3 were obtained numerically.  They are consistent with known analytic results \cite{Adkins93} that are listed in table as ``total'' contributions for the SE, V, and lad corrections.  The NLO energy correction is $\Delta E_{NLO} = I_{NLO} (m \alpha^6/\pi^2)$ where
\be \label{order_alpha}
I_{NLO} = \left ( \frac{21}{16} \zeta(3) - \frac{3}{2} \zeta(2) \ln2 - \frac{13}{6} \zeta(2) -\frac{5}{2} \ln2 + \frac{23}{4} \right ) + i \pi \left ( -\frac{3}{8} \zeta(2) + \frac{5}{4} \right ) .
\ee
This result is consistent with the known $O(\alpha)$ energy correction \cite{Cung78,Adkins93}.  The imaginary parts for the various contributions are related to decay rate contributions as discussed above and are in agreement with the results of \cite{Harris57, Tomozawa80}.

The only NLO contribution that was problematic to evaluate numerically was SE2.  The problem was in the large-$p$ region of the integration space.  The integral was clearly convergent for large $p$ but could only be evaluated to relatively low precision before numerical difficulties set in.  Our solution was to break the large-$p$ region up into smaller pieces and integrate them separately.  Specifically, we used the change of variables $p=x_p/(1-x_p)$, $dp = (1+p)^2 dx_p$ with $0 \le x_p < 1$ to map the infinite $p$ range onto a finite interval.  Then we broke the region $0 \le x_p <1$ up into subintervals 1, 2, 3, $\cdots$ where the $n^{th}$ interval covers $1-s^{n-1} < x_p < 1-s^n$ with $s = \sqrt{1/10}$.  Regions 1 and 2 made the largest contributions but were not problematic to integrate, while the contributions of the higher $n$ regions decreased rapidly with $n$ and were readily integrable to the desired precision.

\begin{table}[t]
\begin{center}
\caption{\label{table1} NLO corrections to the parapositronium energy levels in the $2 \gamma$A channel.  These energy shifts include all multiplicity factors and are given in units of $m \alpha^6/\pi^2$.  The ``IR div'' part should be multiplied by $\ln (\lambda) I_\mathrm{LO}$, and the ``Imaginary Part" must be multiplied by $i \pi$.  The results for SE1, V1, and lad1 were obtained analytically in the current work, while those for SE2, V2, V3, lad2, and lad3 were obtained numerically in the current work.  Analytic results for the total SE, V, and lad contributions were taken from \cite{Adkins93} and are consistent with the numerical values obtained here.}
\begin{ruledtabular}
\begin{tabular}{cccc}
Term & IR div & Real Part & Imaginary Part \\
\hline\noalign{\smallskip}
VP & 0 & $-\frac{1}{6} \zeta(2)$ & 0 \\
\hline
SE1 & 2 & $\half \ln 2 - \half$ & $-\fourth$  \\
SE2 & 0 & -0.092016(4) & $-\ln 2$ \\
SE total & 2 & $-\half \zeta(2) + \ln^2 2 + \half \ln 2 - \fourth$ & $-\ln 2 - \fourth$ \\
\hline
V1 & -4 & $-\frac{5}{2} \ln 2 + \frac{5}{2}$ & $\frac{5}{4}$ \\
V2 & 0 & $0.7628944(2)$ & $-\frac{3}{4} \zeta(2) + \ln 2 + \fourth$ \\
V3 & 0 & $0.5594205(7)$ & $\frac{3}{8} \zeta(2) - \half$ \\
V total & -4 & $\frac{21}{16} \zeta(3) - \frac{3}{2} \zeta(2) \ln2$ & $-\frac{3}{8} \zeta(2) + \ln2 + 1$ \\
& & $- \fourth \zeta(2) - \ln^2 2 - 2 \ln 2 + \frac{9}{2}$ & \\
\hline
lad1 & 2 & $1 - \ln2$ & $\half$ \\
lad2 & 0 & $-0.9739264(4)$ & $0.346573594(6)$ \\
lad3 & 0 & $-0.5822406(3)$ & $-0.346573594(4)$ \\
lad total & 2 & $-\frac{5}{4} \zeta(2) - \ln2 + \frac{3}{2}$ & $\half$ \\
\hline
total & 0 & $\frac{21}{16} \zeta(3) - \frac{3}{2} \zeta(2) \ln2$  &$-\frac{3}{8} \zeta(2) + \frac{5}{4}$ \\
& & $ - \frac{13}{6} \zeta(2) -\frac{5}{2} \ln2 + \frac{23}{4} $ &  \\
\end{tabular}
\end{ruledtabular}
\end{center}
\end{table}


\section{The ``product'' contributions}
\label{product}

\begin{figure}
\includegraphics[width=6.2in]{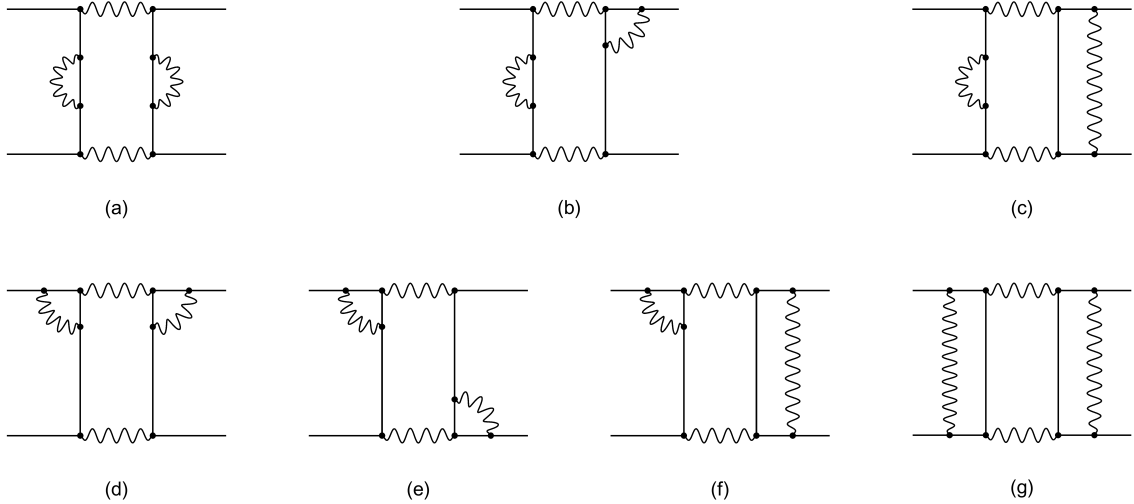}
\caption{\label{fig4} The seven types of NNLO terms in the $2 \gamma \mathrm{A}$ channel having a one-loop correction on each side of the annihilation event.  The contributions are: (a) self energy--self energy, (b) self energy--vertex, (c) self energy--ladder, (d) vertex--vertex (type A), (e) vertex--vertex (type B), (f) vertex--ladder, and (g) ladder--ladder.  Each of these diagrams represents two configurations of annihilation photons (uncrossed and crossed); additional factors have been included to account for the various places in the diagram where the corrections could act.}
\end{figure}

The ``product'' corrections have a one-loop correction--self energy, vertex, or ladder--on each side of the central annihilation event.  These next-to-next-to-leading-order (NNLO) contributions are shown in Fig.~\ref{fig4}.  The symmetry factors associated with these seven graphs are \{2, 8, 4, 4, 4, 8, 2 \} respectively coming from uncrossed and crossed annihilation photons and the possible positions of the one-loop parts in the diagram, all of which contribute equal energy corrections.  Figs.~4(d) and 4(e) represent the two types of vertex--vertex corrections--those acting on the same annihilation photon (``type A'') and those acting on different photons (``type B'').

Numerical results for the various produce corrections are shown in Table~\ref{table2}.  The real parts were the result of straightforward, if sometimes extremely time-consuming, numerical integration using the adaptive Monte Carlo integration routine Vegas \cite{Lepage78}.  All contributions that involved the SE2 correction were done by separation into regions as for the one-loop SE2 contribution described above.  The imaginary parts were done numerically and checked against analytic results obtained by the following considerations.  The imaginary part of the energy correction is related to the decay rate correction according to (\ref{rate_imE}).  There is only one independent amplitude for the decay of pseudoscalar $\ell=0$ parapositronium to two real photons: $\hat k \cdot \hat \epsilon^*_1 \times \epsilon^*_2$, where $\hat k$ represents the direction of the momentum of one of the photons (in the center of mass frame) and $\hat \epsilon_i$ is the polarization vector of the $i^{th}$ photon.  It follows that the amplitude for one-loop corrections to the decay can be written as
\be
A = A_{LO} \left \{ 1+\frac{\alpha}{\pi} \left ( B + C + \cdots \right ) \right \}
\ee
where $A_{LO}$ is the LO decay amplitude and $B$ and $C$ represent one-loop corrections.  Then the decay rate, as corrected by $B$ and $C$ (coming from SE2 and V2 for example), is
\be \label{rate_expansion}
\Gamma = \Gamma_{LO} \left \{ 1 + \frac{2 \alpha}{\pi} \left ( B + C + \cdots \right ) + \frac{\alpha^2}{\pi^2} \left ( B + C + \cdots \right )^2 \right \}
\ee
where $\Gamma_{LO} = m \alpha^5/2$ is the LO rate.  On conversion into an expression for the imaginary part of the energy correction, (\ref{rate_expansion}) becomes
\be
\mathrm{Im} (\Delta E) = \Bigl ( -\frac{\pi}{4} \Bigr ) \frac{m \alpha^5}{\pi} + \Bigl ( -\frac{\pi}{2} \left (B+C+\cdots \right ) \Bigr ) \frac{m \alpha^6}{\pi^2} + \Bigl ( - \frac{\pi}{4} \left ( B+C+\cdots \right )^2 \Bigr ) \frac{m \alpha^7}{\pi^3} + \cdots
\ee
so that, for the one-loop correction due to $B$, for instance,
\be
\mathrm{Im} (I_B) = -\frac{\pi}{2} B,
\ee
while for two-loop corrections with a $B$ correction on each side one has
\be
\mathrm{Im} (I_{BB}) = -\frac{\pi}{4} B^2 = -\frac{1}{\pi} \mathrm{Im} (I_B)^2
\ee
and for a two-loop correction with a $B$ on one side and a $C$ on the other, one has
\be
\mathrm{Im} (I_{BC}) = -\frac{\pi}{2} B C = -\frac{2}{\pi} \mathrm{Im} (I_B) \mathrm{Im} (I_C) .
\ee
So all NNLO imaginary parts of the product type can be expressed analytically in terms of the known NLO imaginary parts.  The numerical results we obtained for the NNLO imaginary parts are in accord with these analytic results.  The total imaginary part from the contributions listed in Table~\ref{table2} is $-\pi (\pi^2+4)^2/256$.

\begin{table}
\begin{center}
\caption{\label{table2} Contributions to the p-Ps energy levels coming from products of one-loop corrections, one on either side of the annihilation event.  The one-loop parts are the self energy (SE2), vertex (V2 and V3), and ladder (lad2 and lad3) corrections.  For type A vertex corrections the two vertex parts act on the same photon, while for type B corrections they act on different photons.  The energy shifts are given in units of $m \alpha^7/\pi^3$. }
\begin{ruledtabular}
\begin{tabular}{cccc}
Term & Figure & Real Part & Imaginary Part \\
\hline\noalign{\smallskip}
SE2-SE2 & 3a & 0.188882(5) & -1.509388  \\
SE2-V2 & 3b & 1.384328(5) & -1.265410 \\
SE2-V3 & 3b & 0.700844(8) & 0.508903 \\
SE2-lad2 & 3c & -1.795766(8) & 1.509388 \\
SE2-lad3 & 3c & -0.582731(4) & -1.509388 \\
V2-V2 (A) & 3d & 1.024802(6) & -0.132609 \\
V2-V3 (A) & 3d & 0.056872(3) & 0.106661 \\
V3-V3 (A) & 3d & -0.113306(3) & -0.021448 \\
V2-V2 (B) & 3e & -0.344233(4) & -0.132609 \\
V2-V3 (B) & 3e & -0.079219(3) & 0.106661 \\
V3-V3 (B) & 3e & -0.117118(3) & -0.021448 \\
V2-lad2 & 3f & -1.854200(8) & 0.632705 \\
V2-lad3 & 3f & 0.379183(3) & -0.632705 \\
V3-lad2 & 3f & -0.063542(8) & -0.254452 \\
V3-lad3 & 3f & 0.609061(8) & 0.254452 \\
lad2-lad2 & 3g & 1.241728(7) & -0.377347 \\
lad2-lad3 & 3g & -0.422514(3) & 0.754694 \\
lad3-lad3 & 3g & -0.371805(8) & -0.377347 \\
\hline
total & & -0.158734(25) & -2.360685 \\
\end{tabular}
\end{ruledtabular}
\end{center}
\end{table}

We can now work out the total NNLO correction to the parapositronium energy levels coming from product contributions.  It is useful to write the NLO corrections as
\bse \bearray
I_{SE} &=& 2 S_1 I_\mathrm{LO} + I_\mathrm{SE2} , \\
I_{V} &=& 4 V_1 I_\mathrm{LO} + I_\mathrm{V23} , \\
I_\mathrm{lad} &=& 2 L_1 I_\mathrm{LO} + I_\mathrm{lad23}
\eearray \ese
where $S_1$ is given in (\ref{S1_def}), $V_1 = -\ln \lambda - 5/4$, $L_1$ is given in (\ref{L1_def}), and
\bse \bearray
I_\mathrm{SE2} &=& \Bigl ( -\half \zeta(2) + \ln^2 2 + \fourth \Bigr ) + i \pi \Bigl ( -\ln 2 \Bigr ) , \\
I_\mathrm{V23} &=& \Bigl ( \frac{21}{16} \zeta(3) - \frac{3}{2} \zeta(2) \ln 2 - \fourth \zeta(2) - \ln^2 2 + \half \ln 2 + 2 \Bigr ) + i \pi \Bigl ( -\frac{3}{8} \zeta(2) + \ln 2 - \fourth \Bigr ) , \\
I_\mathrm{lad23} &=& \Bigl ( -\frac{5}{4} \zeta(2) + \half \Bigr ) .
\eearray \ese
Finally, the NNLO contribution from all ``product'' contributions is given by the sum of the following six terms
\bse \label{product_contributions} \bearray 
I_\mathrm{SE-SE} &=& S_1^2 I_\mathrm{LO} + S_1 I_\mathrm{SE2} + I_\mathrm{SE2-SE2} , \\
I_\mathrm{SE-V} &=& 4 S_1 V_1 I_\mathrm{LO} + S_1 I_\mathrm{V23} + 2 V_1 I_\mathrm{SE2} + I_\mathrm{SE2-V23} , \\
I_\mathrm{SE-lad} &=& 2 S_1 L_1 I_\mathrm{LO} + S_1 I_\mathrm{lad23} + L_1 I_\mathrm{SE2} + I_\mathrm{SE2-lad23} , \\
I_\mathrm{V-V} &=& 4 V_1^2 I_\mathrm{LO} + 2 V_1 I_\mathrm{V23} + I_\mathrm{V23-V23} , \\
I_\mathrm{V-lad} &=& 4 V_1 L_1 I_\mathrm{LO} + L_1 I_\mathrm{V23} + 2 V_1 I_\mathrm{lad23} + I_\mathrm{V23-lad23} , \\
I_\mathrm{lad-lad} &=& L_1^2 I_\mathrm{LO} + L_1 I_\mathrm{lad23} + I_\mathrm{lad23-lad23} ,
\eearray \ese
where $I_\mathrm{SE2-SE2}$, etc., are taken from Table~\ref{table2}, $V23$ and $\mathrm{lad}23$ include both parts 2 and 3, and $I_\mathrm{V-V}$ is the sum of the six $\mathrm{V-V}$ contributions including both parts A and B.  Upon adding up the contributions of (\ref{product_contributions}) we obtain the total product contribution
\be
I_\mathrm{product} = 9 I_\mathrm{LO} - 3 \left ( I_\mathrm{SE2} + I_\mathrm{V23} + I_\mathrm{lad23} \right ) + I_\mathrm{XY} 
\ee
where we have used
\be
S_1 + 2 V_1 + L_1 = -3 
\ee
and $I_\mathrm{XY}$ is the total from Table~\ref{table2}:
\be
I_\mathrm{XY} = -0.158734(25) - i \pi \Bigl ( \frac{\pi^2}{16} + \fourth \Bigr )^2 .
\ee
In all, then, the final ``product'' result is
\be \label{result}
I_\mathrm{product} = -0.561971(25) - i \pi \Bigl ( \frac{\pi^2}{16} - \frac{5}{4} \Bigr )^2 .
\ee


\section{Results}
\label{results}

On combining the light-by-light contribution of Fig.~1(a) \cite{Adkins14b} and the vacuum polarization contribution of Fig.~1(b) \cite{Adkins15a} with the new product contribution of Fig.~1(c) given in \eqref{result}, we have for the NNLO $2 \gamma \mathrm{A}$ energy corrections the results
\bse \bearray
\Delta E_{\mathrm{LbyL}} &=& \Bigl \{ 1.58377(8) - 1.016262(15) i \Bigr \} \frac{m \alpha^7}{\pi^3}, \\
\Delta E_{\mathrm{vac. pol.}} &=& \Bigl \{ -0.153095(3) \Bigr \} \frac{m \alpha^7}{\pi^3}, \\
\Delta E_{\mathrm{product}} &=& \Bigl \{ -0.561971(25) - 1.259397 i \Bigr \} \frac{m \alpha^7}{\pi^3},
\eearray \ese
with the NNLO contribution of Fig.~1(d) not yet known.  These make a net contribution to the real part of the parapositronium energy levels of
\be \label{net_real}
\mathrm{Re} \bigl ( \Delta E_{\mathrm{net}} \bigr ) = 0.86870(9) \frac{m \alpha^7}{\pi^3} = 3.81 kHz.
\ee
Numerically, this is significantly below the present experimental precision of order $1 MHz$.  Expression  (\ref{net_real}) applies to all $S$ states, although an additional factor of $1/n^3$ must be applied to states with higher principal quantum number $n$.

Our results are also applicable to true muonium, the $\mu^+ \mu^-$ bound state.  Even though true muonium has not yet been detected it has interesting properties and is the subject of current searches.  Some recent discussions with references to earlier works include Refs.~\cite{Jentschura97,Brodsky09,Banburski12,Chliapnikov14,Ellis15,Lamm15}.  All of the positronium corrections considered in this paper apply as well to true muonium with the electron mass $m$ replaced by the muon mass $m_\mu \sim206.8 m$.  The net energy shift for true muonium from known $2 \gamma \mathrm{A}$ contributions, the analog of \eqref{net_real}, is
\be
\mathrm{Re} \bigl ( \Delta E_{\mathrm{net}}^{\mu^+ \mu^-} \bigr ) = 0.86870(9) \frac{m_\mu \alpha^7}{\pi^3} = 0.79 MHz.
\ee
This results takes into account light-by-light and vacuum polarization contributions with muons both in the bound state and in the virtual fermion loops.  Additional contributions analogous to the light-by-light and vacuum polarization ones but with electrons in the fermion loops instead of muons also exist but aren't calculated here.

In summary, we have obtained the third of four classes of NNLO $2 \gamma \mathrm{A}$ contributions to the parapositronium energies at order $m \alpha^7$ and the analogous corrections to the energies of true muonium.


\begin{acknowledgments}
We acknowledge the support of the National Science Foundation through Grant No. PHY-1404268 and of the Franklin \& Marshall College Grants Committee through the Hackman Scholars Program.
\end{acknowledgments}


     



\begin{thebibliography}{10}

\bibitem{Deutsch51a} M. Deutsch, Phys. Rev. {\bf 82}, 455 (1951).
\bibitem{Deutsch51b} M. Deutsch and E. Dulit, Phys. Rev. {\bf 84}, 601 (1951).
\bibitem{Maglic75} ``Discovery of Positronium'', in {\it Adventures in Experimental Physics}, vol. 4, edited by B. Maglic (World Science Education, Princeton, NJ, 1975), pp. 64-127.
\bibitem{Debenedetti54} S. DeBenedetti and H. C. Corben, Ann. Rev. Nucl. Sci. {\bf 4}, 191 (1954).
\bibitem{Hughes73} V.W. Hughes, in Physik 1973, Plenarvorträge Physikertagung, 37th (Physik Verlag, Weinheim, Germany), p. 123, 1973.
\bibitem{Berko80} S. Berko and H.N. Pendleton, Ann. Rev. Nucl. Part. Sci. {\bf 30}, 543 (1980).
\bibitem{Rich81} A. Rich, Rev. Mod. Phys. {\bf 53}, 127 (1951).
\bibitem{Mills90} A.P. Mills, Jr., and S. Chu, in {\it Quantum Electrodynamics}, edited by T. Kinoshita (World Scientific, Singapore), pp. 774-821, 1990.
\bibitem{Karshenboim04} S.G. Karshenboim, Int. J. Mod. Phys. A {\bf 19}, 3879 (2004).
\bibitem{Karshenboim05} S.G. Karshenboim, Phys. Rep. {\bf 422}, 1 (2005).
\bibitem{Namba12} T. Namba, Prog. Theor. Exp. Phys. {\bf 2012}, 04D003.
\bibitem{Mills75} A. P. Mills, Jr. and G. H. Bearman, Phys. Rev. Lett. {\bf 34}, 246 (1975).
\bibitem{Mills83} A. P. Mills, Jr., Phys. Rev. A {\bf 27}, 262 (1983).
\bibitem{Ritter84} M. W. Ritter, P. O. Egan, V. W. Hughes, and K. A. Woodle, Phys. Rev. A {\bf 30}, 1331 (1984).
\bibitem{Ishida14} A. Ishida {\it et al.}, Phys. Lett. B {\bf 734}, 338 (2014).
\bibitem{Mills75b} A. P. Mills, Jr., S. Berko, and K. F. Canter, Phys. Rev. Lett. {\bf 34}, 1541 (1975).
\bibitem{Hatamian87} S. Hatamian, R. S. Conti, and A. Rich, Phys. Rev. Lett. {\bf 58}, 1833 (1987).
\bibitem{Hagena93} D. Hagena, R. Ley, D. Weil, G. Werth, W. Arnold, and H. Schneider, Phys. Rev. Lett. {\bf 71}, 2887 (1993).
\bibitem{Ley94} R. Ley, D. Hagena, D. Weil, G. Werth, W. Arnold, and H. Schneider, Hyperfine Interact. {\bf 89}, 327 (1994).
\bibitem{Fee93} M. S. Fee, A. P. Mills, Jr., S. Chu, E. D. Shaw, K. Danzmann, R. J. Chichester, and D. M. Zuckerman, Phys. Rev. Lett, {\bf 70}, 1397 (1993).

\bibitem{Fan96} S. Fan, C. D. Beling, and S. Fung, Phys. Lett. A {\bf 216}, 129 (1996).
\bibitem{Ley02} R. Ley, Appl. Surf. Sci. {\bf 194}, 301 (2002).
\bibitem{Cassidy08} D. B. Cassidy, H. W. K. Tom, and A. P. Mills, Jr., AIP Conf. Proc. {\bf 1037}, 66 (2008).
\bibitem{Sasaki11} Y. Sasaki {\it et al.}, Phys. Lett. B {\bf 697}, 121 (2011).
\bibitem{Crivelli11} P. Crivelli, C. L. Cesar, and U. Gendotti, Can. J. Phys. {\bf 89}, 29 (2011).
\bibitem{Ishida12} A. Ishida {\it et al.} Hyperfine Interact. {\bf 212}, 133 (2012).
\bibitem{Yamazaki12} T. Yamazaki {\it et al.}, Phys. Rev. Lett. {\bf 108}, 253401 (2012).
\bibitem{Cassidy12} D. B. Cassidy, T. H. Hisakado, H. W. K. Tom, and A. P. Mills, Jr., Phys. Rev. Lett. {\bf 109}, 073401 (2012).
\bibitem{Mills14} A. P. Mills, Jr., J. Phys.: Conf. Ser. {\bf 488}, 012001 (2014).
\bibitem{Cooke15} D. A. Cooke, P. Crivelli, J. Alnis, A. Antognini, B. Brown, S. Friedreich, A. Gabard, T. W. Haensch, K. Kirch, A. Rubbia, and V. Vrankovic, Hyperfine Interact. (March, 2015).
\bibitem{Miyazaki15} A. Miyazaki, T. Yamazaki, T. Suehara, T. Namba, S. Asai, T. Kobayashi, H. Saito, Y. Tatematsu, I. Ogawa, and T. Idehara, Prog. Theor. Exp. Phys. 011C01 (2015).
\bibitem{Elkhovsky94} A. S. Elkhovsky, I. B. Khriplovich, and A. I. Mil'stein, Zh. Eksp. Teor. Fiz. {\bf 105}, 299 (1994) [Sov. Phys. JETP {\bf 78}, 159 (1994)].
\bibitem{Pachucki98} K. Pachucki and S. G. Karshenboim, Phys. Rev. Lett. {\bf 80}, 2101 (1998).
\bibitem{Czarnecki99a} A. Czarnecki, K. Melnikov, and A. Yelkhovsky, Phys. Rev. Lett. {\bf 82}, 311 (1999); Phys. Rev. A {\bf 59}, 4316 (1999).
\bibitem{Zatorski08} J. Zatorski, Phys. Rev. A {\bf 78}, 032103 (2008).
\bibitem{Karshenboim93} S. G. Karshenbo\u\i m, Zh. Eksp. Teor. Fiz. {\bf 103}, 1105 (1993) [Sov. Phys. JETP {\bf 76}, 541 (1993)].
\bibitem{Melnikov99} K. Melnikov and A. Yelkhovsky, Phys. Lett. B {\bf 458}, 143 (1999).
\bibitem{Pachucki99} K. Pachucki and S. G. Karshenboim, Phys. Rev. A {\bf 60}, 2792 (1999).
\bibitem{Kniehl00} B. A. Kniehl and A. A. Penin, Phys. Rev. Lett. {\bf 85}, 5094 (2000).
\bibitem{Melnikov01} K. Melnikov and A. Yelkhovsky, Phys. Rev. Lett. {\bf 86}, 1498 (2001).
\bibitem{Hill01} R. J. Hill, Phys. Rev. Lett. {\bf 86}, 3280 (2001).
\bibitem{Marcu11} S. R. Marcu, Master's thesis, University of Alberta, 2011.
\bibitem{Baker14} M. Baker, P. Marquard, A. A. Penin, J. Piclum, and M. Steinhauser, Phys. Rev. Lett. {\bf 112}, 120407 (2014).
\bibitem{Adkins14a} G. S. Adkins and R. N. Fell, Phys. Rev. A {\bf 89}, 052518 (2014).
\bibitem{Eides1415} M. I. Eides and V. A. Shelyuto, Phys. Rev. D {\bf 89}, 111301(R) (2014); {\bf 92}, 013010 (2015).
\bibitem{Adkins15c} G. S. Adkins, M. Kim, C. Parsons, and R. N. Fell, Phys. Rev. Lett. {\bf 115}, 233401 (2015).
\bibitem{Adkins14b} G. S. Adkins, C. Parsons, M. D. Salinger, R. Wang, and R. N. Fell, Phys. Rev. A {\bf 90}, 042502 (2014).
\bibitem{Adkins15a} G. S. Adkins, C. Parsons, M. D. Salinger, and R. Wang, Phys. Lett. B {\bf 747}, 551 (2015).

\bibitem{Karplus52} R. Karplus and A. Klein, Phys. Rev. {\bf 87}, 848 (1952).
\bibitem{Cung78} V. K. Cung, A. Devoto, T. Fulton, and W. W. Repko, Phys. Lett. B {\bf 78}, 116 (1978); Phys. Rev. A {\bf 19}, 1886 (1979).
\bibitem{Adkins93} G. S. Adkins, Y. M. Aksu, and M. H. T. Bui, Phys. Rev. A {\bf 47}, 2640 (1993).
\bibitem{Adkins99} G. S. Adkins and R. N. Fell, Phys. Rev. A {\bf 60}, 4461 (1999).
\bibitem{Itzykson80} We use the metric and gamma matrix conventions of C. Itzykson and J.-B. Zuber, {\it Quantum Field Theory\/} (McGraw-Hill, New York, 1980).
\bibitem{Cutkosky60} R. E. Cutkosky, J. Math. Phys. {\bf 1}, 429 (1960).
\bibitem{Adkins01a} G. S. Adkins, R. N. Fell, and J. Sapirstein, Phys. Rev. A {\bf 63}, 032511 (2001).

\bibitem{Adkins01b} G. S. Adkins, R. N. Fell, and J. Sapirstein, Phys. Rev. D {\bf 63}, 125009 (2001).
\bibitem{Adkins02} G. S. Adkins, R. N. Fell, and J. Sapirstein, Ann. Phys. (N.Y.) {\bf 295}, 136 (2002).
\bibitem{Harris57} I. Harris and L. M. Brown, Phys. Rev. {\bf 105}, 1656 (1957).
\bibitem{Alekseev59} A. I. Alekseev, Zh. Eksp. Teor. Fiz. {\bf 36}, 1437 (1959) [Sov. Phys. JETP {\bf 9}, 1020 (1959)].
\bibitem{Stroscio74} J. M. Stroscio and J. M. Holt, Phys. Rev. A {\bf 10}, 749 (1974).
\bibitem{Caswell77} W. E. Caswell, G. P. Lepage, and J. Sapirstein, Phys. Rev. Lett. {\bf 38}, 488 (1977).
\bibitem{Caswell86} W. E. Caswell and G. P. Lepage, Phys. Lett. B {\bf 167}, 437 (1986).
\bibitem{Labelle94} P. Labelle, Ph.D. thesis, Cornell University, 1994.
\bibitem{Tomozawa80} Y. Tomozawa, Ann. Phys. (N.Y.) {\bf 128}, 463 (1980).
\bibitem{Lepage78} G. P. Lepage, J. Comput. Phys. {\bf 27}, 192 (1978).

\bibitem{Jentschura97} U. D. Jentschura, G. Soff, V. G. Ivanov, and S. G. Karshenboim, Phys. Rev. A {\bf 56}, 4483 (1997).
\bibitem{Brodsky09} S. J. Brodsky and R. F. Lebed, Phys. Rev. Lett. {\bf 102}, 213401 (2009).
\bibitem{Banburski12} A. Banburski and P. Schuster, Phys. Rev. D {\bf 86}, 093007 (2012).
\bibitem{Chliapnikov14} P. V. Chliapnikov, Report No. DIRAC-NOTE-2014-05 (2014).
\bibitem{Ellis15} S. C. Ellis and J. Bland-Hawthorn, Phys. Rev. D {\bf 91}, 123004 (2015).
\bibitem{Lamm15} H. Lamm, Phys. Rev. D {\bf 91}, 073008 (2015).


\end{thebibliography}
\end{document}